\documentclass[aps,twocolumn,superscriptaddress]{revtex4-1}
\usepackage{import}
\usepackage[colorinlistoftodos]{todonotes}
\newcommand{\proj}{\hat{\mathcal{P}}_{T_{\ket{\psi{(\{A\})}}}}}

\usepackage[utf8]{inputenc}

\usepackage[english]{babel} 
\usepackage{amsmath}
\usepackage{float}
\usepackage{listings}
\usepackage{amsfonts}
\usepackage{amssymb}
\usepackage{braket}
\usepackage{graphicx}
\usepackage{color}
\usepackage{calc}
\usepackage[left=2cm,right=2cm,top=2cm,bottom=2cm]{geometry}
\usepackage{pgfplots}
\usepackage{tikz}
\usepackage[export]{adjustbox}
\usepackage{amsmath}

\pgfplotsset{width=15cm,compat=1.9}

\DeclareMathOperator{\Tr}{Tr}

\graphicspath{{./figures/}}

\newcommand{\inc}[2][0.0]{\raisebox{#1ex}{\includegraphics[valign=c]{#2.pdf}}}

\newcommand{\eq}[1]{eq.~\ref{eq:#1}}
\newcommand{\Eq}[1]{Eq.~\ref{eq:#1}}
\newcommand{\fig}[1]{fig.~\ref{fig:#1}}

\newcommand{\eqs}[2]{eqs \ref{eq:#1} and \ref{eq:#2}}

\newcommand{\etal}{\textit{et al}}

\newcommand{\executeiffilenewer}[3]{%
\ifnum\pdfstrcmp{\pdffilemoddate{#1}}%
{\pdffilemoddate{#2}}>0%
{\immediate\write18{#3}}\fi%
}

\begin{document}
\title{Dynamically Evolving Bond-Dimensions within the one-site Time-Dependent-Variational-Principle method for Matrix Product States: Towards efficient simulation of non-equilibrium open quantum dynamics }
\author{Angus J. Dunnett}
\author{Alex W. Chin}
\affiliation{Sorbonne Universit\'{e}, CNRS, Institut des NanoSciences de Paris, 4 place Jussieu, 75005 Paris, France}

\begin{abstract}
Understanding the emergent system-bath correlations in non-Markovian and non-perturbative open systems is a theoretical challenge that has benefited greatly from the application of Matrix Product State (MPS) methods. Here, we propose an autonmously adapative variant of the one-site Time-Dependent-Variational-Principle (1TDVP) method for  many-body MPS wave-functions in which the local bond-dimensions can evolve to capture growing
  entanglement 'on the fly'. We achieve this by efficiently examining the effect of increasing each MPS bond-dimension in advance of each dynamic timestep, resulting in an MPS that can dynamically and inhomogeneously restructure itself as the complexity of the dynamics grows across time and space. This naturally leads to more efficient simulations, oviates the need for multiple convergence runs, and, as we demonstrate, is ideally suited to the typical, finite-temperature 'impurity' problems that describe open quantum system connected to multiple environments.  
\end{abstract}
\maketitle

\section{Introduction}
\label{sec:intro}

\noindent
The emergent irreversible phenomena of thermalisation, decoherence and transport appear ubiquitously in quantum devices and
critically determine how physical, molecular - and even biological - processes are able to exploit, capture or convert
energy on the nanoscale \cite{weiss2012quantum,breuer2002theory,chin2013role,bredas2017photovoltaic,scholes2017using}. However, perturbative master equation approaches to 'open' quantum systems fail in the presence of strong
system-bath coupling and/or non-equilibrium environments \cite{weiss2012quantum}. What then emerges
is a situation in which strong and time-evolving correlations arise between the system and its environmental excitations. The boundary between system and environment becomes blurred and transitory, and the key
physics can only be understood by describing the joint quantum dynamics of the system and the environment on an
essentially equal footing. Unfortunately, given that typical environments contain a continuum of excitation modes, this quantum many body problem
would appear to be a daunting computational challenge \cite{ashida2020quantum}.

Two major approaches to this problem have emerged over recent years: one branch aims to efficiently simulate the
propagators of the system's reduced density matrix \cite{weiss2012quantum,strathearn_efficient_2018,ishizaki2009unified,topaler1994quantum}, the other - which we shall pursue - aims at representing and
evolving the entire system-environment wave function. Important contributions in this latter domain are the DMRG-based TEDOPA
technique \cite{prior2010efficient, PhysRevA.93.020102}, Dissipation-Assisted Matrix Product Factorization \cite{PhysRevLett.123.100502}, Time-Dependent Numerical Renormalisation Group techniques and the Multi-Layer Multi Configurational
Time-Dependent Hartree method (ML-MCTDH) developed in Chemical physics \cite{lindner2019renormalization, wang2019quantum}. In this contribution, we present a powerful and versatile extension of a variational technique connected to the simulation of Tensor Network dynamics for open system models\cite{haegeman_unifying_2016,haegeman_time-dependent_2011,PhysRevB.93.075105,PhysRevB.96.115427}. 

The Time-Dependent-Variational-Principle (TDVP), formulated for Matrix-Product-States (MPS)
\cite{orus_practical_2014,schollwock_density-matrix_2011,paeckel_time-evolution_2019} by Haegeman \etal
\cite{haegeman_unifying_2016,haegeman_time-dependent_2011}, has established itself as a powerful method for time evolving
large, many body wave functions of both 1D and quasi-1D (ie. tree structured) systems \cite{bauernfeind_time_2020}. Its
chief advantages over other methods being the simplicity with which long-range interactions can be handled
\cite{hubig_generic_2017,li_generalization_2019,murg_matrix_2010}, its parallelization \cite{secular_parallel_2020} and its ability to accurately predict the long-time
thermalization dynamics of local observables using MPS of small bond dimension
\cite{leviatan_quantum_2017,goto_performance_2019}. The latter is a consequence of the particular fashion in which the
bond-dimension $ D$ of the evolved MPS is maintained, and is particularly advantageous for the simulation of open systems, where a centralised (local) quantum system is coupled to an extended environment.  In contrast to other methods such as TEBD
\cite{vidal_efficient_2004,daley_time-dependent_2004,white_real-time_2004} that uses a truncation after each step to
prevent the bond-dimension growing indefinitely, TDVP attempts to find the MPS with \emph{fixed bond-dimensions} $D$
which best approximates the time-evolved wave-function $\ket{\Psi(t)}$. This variational problem, which was originally considered by
Dirac and Frenkel \cite{heller1976time}, can be solved by applying an
orthogonal projector $\proj$ onto the tangent space of $\ket{\psi(\{A\})}$ with bond-dimension $D$. This gives rise to
the following equation for a wave-function $\ket{\Psi}$ evolving under an Hamiltonian $\hat{H}$ 
\begin{equation}
  \label{eq:tdvp}
  \frac{d}{dt}\ket{\Psi(t,\{A\})} = -i\proj\hat{H}\ket{\Psi(t,\{A\})}.
\end{equation}
The introduction of the projector into the Schroedinger equation, of course, introduces an error, known as the
projection error \cite{hubig_error_2018}. This is due to the fact that $\ket{\Psi(t)}$ will in general become highly entangled, and so cannot be
represented by an MPS ansatz $\Psi(t,\{A\}$ in which the entanglement that can be accommodated is bounded by $\log(D)$. Crucially however, the projection
error will not lead to an error in the energy, nor in any of the other quantities conserved by the Hamiltonian. The projection error will also leave the normalisation of the state unchanged \cite{lubich_time_2015,haegeman_unifying_2016}. 

Using a formally exact splitting of the projector \cite{lubich_time_2015} \Eq{tdvp} can be solved numerically by a series of
local updates on the site tensors using effective Hamiltonians. Each local update preserves the bond-dimensions of each
site. The bond-dimension thus remains constant through the simulation.

However, the fixed nature of the bond-dimension in TDVP can be disadvantageous in certain
situations. First, in many cases, having a fixed bond-dimension is highly sub-optimal. Often one is interested in
following the dynamics of an initial state with an exact low bond-dimension MPS representation, most commonly a product
state with trivial bond-dimension. Time evolution will inevitably lead to the growth of entanglement between sites which
will require a larger bond-dimension to capture. In practice this means that to track the dynamics to long times one has
to embed the initial, low-entanglement MPS, in a manifold of larger bond-dimension (or perform some other
time-evolution method such as 2TDVP or global Krylov for a short time) and hope that the bond-dimension chosen is
sufficient to capture the entanglement which will develop. Thus, at short times, the bond-dimension will inevitably be
far superior to what is required and as the complexity of TDVP scales with $D^{2}$ it is important to minimize the
bond-dimension wherever possible - especially in tree tensor networks \cite{schroder_tensor_2019}. Furthermore, the required bond-dimension will often be strongly dependent on the
physical parameters of the simulation and it is difficult to predict or even estimate the one from the other. As a
result one is forced to run multiple simulations at different bond-dimensions and look for convergence in observables,
and this must be done for every physical configuration one wishes to consider. As we shall also see, in problem related to impurities or open systems, the numerical resources ( bond dimensions) grow across the environment within a spreading 'light cone' around the system, and thus using a fixed bond dimension for every site, including the most 'distance' ( vide infra) parts environment, will always very inefficient.   

By analogy to the 'state enrichment' methods for finding ground states in DMRG, a 2-site variant of TDVP (2TDVP) has been also been recently developed. Here, instead of local updates on single sites, one updates pairs of neighboring sites together\cite{xie_time-dependent_2019}. These pair updates have
the effect of taking the MPS out of its initial manifold by increasing the bond-dimension. To stop the bond-dimension
growing indefinitely, it is then necessary to perform a truncation. The error associated with these truncations is known as
the truncation error \cite{schollwock_density-matrix_2011}. The truncation error can by controlled by changing the threshold for truncation, allowing a
kind of dynamic optimization of the bond-dimensions; where there is high entanglement between sites, the bond-dimension will become
large as the singular values will decay less quickly, and where the entanglement is low the bond-dimension can be
small, thus saving computational resources. However the new truncation error introduced in 2TDVP does not conserve the norm or energy of the evolving state, involves costly full SVDs and loses the attractive geometrically guaranteed properties of the tangent space approach.

In this article we will attempt to combine the advantages of both approaches by introducing an efficient evaluation of
the projection error of one-site TDVP (1TDVP) to dynamically optimize the \emph{local} bond-dimensions 'on the fly', that is to say, during
the course of a single run of the proposed 1TDVP algorithm. This will allow us to track changes in entanglement during
the evolution without any prior knowledge, and creates a bond-adaptive MPS \emph{structure} that evolves to handle
emerging quantum correlations while avoiding truncation errors completely. As we we shall later show, such a capability
leads to significant gains in time and computing resources over standard 1TDVP in non -equilibrium open system problems
- and also provides direct, real-time insight into the emergence and spread of the correlations that drive the emergence of dissipative local dynamics.

The structure of this article is as follows. In section \ref{sec:sse} we will give a brief overview of MPS and introduce
the notion of 1-site subspace expansions \cite{hubig_strictly_2015}. In section \ref{sec:tdvp} we will
develop the idea of sub-space expansion into a way of dynamically increasing the bond-dimensions in 1TDVP. In section
\ref{sec:eg} we will apply this to demonstrate the advantages of our method over the standard fixed bond-dimension 1TDVP
in a model of disipative quantum heat flow. Specifically, we shall explore the dynamics resulting from the connection of
a hot and cold bosonic reservoir via a single qubit-like object, and show how our bond-adaptive MPS is ideally suited to
treat the time-varying computational demands required by recent proposals to describe mixed system-bath thermal states
with \emph{pure} wave function dynamics \cite{tamascelli_efficient_2019}.

\section{MPS and sub-space expansion}
\label{sec:sse}

\noindent
The wave-function is written as an MPS with open boundary conditions made up of the set of tensors $\{A\}$ with local
Hilbert space dimensions $\{d\}$ and bond-dimensions $\{D\}$

\begin{equation}
  \label{eq:mps}
  \ket{\psi(\{A\})} = \inc[2]{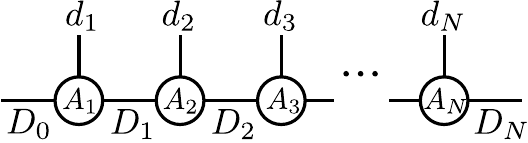}.
\end{equation}
\noindent
The first and final bond-dimension are trivial ($D_{0}=D_{N}=1$) such that contracting the entire network, for a choice of
physical states, will yield a scalar. Similarly, the Hamiltonian is represented as an MPO
\begin{equation}
  \label{eq:hmpo}
  \hat{H} = \inc{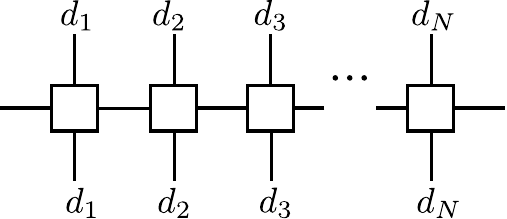}.
\end{equation}

We choose to write our MPS using the convention that a physical leg
pointing downwards implies the elements are complex conjugated. Thus the bra is represented as

\begin{equation}
  \label{eq:bra}
  \bra{\psi(\{A\})} = \inc[-1.4]{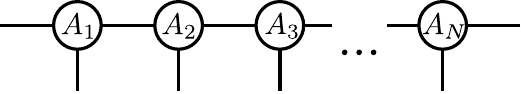}.
\end{equation}
\noindent
One of the key properties of the MPS representation is its gauge freedom which allows different sets of tensors to
represent the same physical state. For example, the transformation $A_{i}\rightarrow A_{i}X$,
$A_{i+1}\rightarrow X^{-1}A_{i+1}$ for any non-singular matrix $X$, leaves the state unchanged. In particular, by
performing QR factorizations, one can put the MPS into the so called canonical forms which are the basis of many MPS
based algorithms.

The QR factorization takes an $m\times n$ rectangular matrix $A$, where $m \ge n$, and decomposes it into an $m\times m$
unitary matrix $Q$ and an $m\times n$ upper triangular matrix $R$. Since $R$ is upper triangular, its bottom $m-n$ rows
are all zero, leading to the following block structure
\begin{equation}
  \label{eq:qr}
  A = QR =
  \begin{pmatrix}
    Q_{1}& Q_{2}
  \end{pmatrix}
  \begin{pmatrix}
    R_{1} \\
    0
  \end{pmatrix}
  = Q_{1}R_{1}.
\end{equation}
\noindent
The matrix $Q_{2}$ consists of $m-n$ orthonormal columns which are orthogonal to the $n$ columns of $Q_{1}$. Since on
multiplying together the factors $Q$ and $R$ the block $Q_{2}$ will simply meet the zero rows of $R$, this block is often
discarded and the factorization is taken as $Q_{1}R_{1}$. This is known as the \textit{thin} or \textit{reduced} QR
factorization whereas taking $QR$ is known as \textit{full} QR. It should be noted that while $Q_{1}$ is unique (provided
that $A$ is full ranked), $Q_{2}$ is not.

Applying this factorization to the tensors in our MPS allows us to decompose $A$ as

\begin{equation}
  \label{eq:leftfac}
  \inc{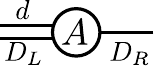} = \inc[1.3]{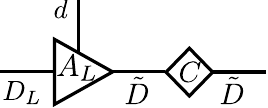},
\end{equation}
\noindent
where $A_{L}$ has the property
\begin{equation}
  \label{eq:idleft}
  \inc{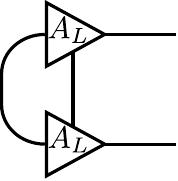} = \hat{1}.
\end{equation}
\noindent
We have written the right bond-dimension of $A_{L}$ above as $\tilde{D}$ to include the possibility of including the one
or more of the columns of $Q_{2}$. We may take $\tilde{D}$ to be any value between $D_{R}$ and $dD_{L}$ inclusive. Taking
$\tilde{D}=D_{R}$ would correspond to \textit{thin} QR, while taking $\tilde{D}=dD_{L}$ would correspond to a
\textit{full} QR.

It is useful to consider the matrix $A$ as a basis transformation whose job is to take the combined Hilbert space of the
$D_{L}$ states from its left plus the $d$ local states and to find the $D_{R}$ most relevant states (where often
$D_{R} \ll dD_{L}$) which it then outputs to the next tensor on the chain. This is how the MPS is able to describe
many-body quantum states using a computationally viable number of parameters. Including the extra states of $Q_{2}$ can
be considered as completing the truncated basis of $D_{R}$ states such that $A$ outputs either a full basis for the
$dD_{L}$ dimensional Hilbert space, or a less severely truncated one. Of course, for a given MPS, including the extra
$Q_{2}$ states will make no difference since the next tensor along the chain will still only except $D_{R}$ states from
its left. However, this notion of sub-space expansion will become important when we come to consider time evolution.

We can equally take the mirror image of \eq{rightfac} and decompose $A$ as
\begin{equation}
  \label{eq:rightfac}
  \inc{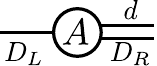} = \inc[1.4]{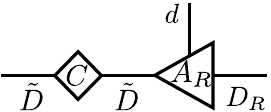},
\end{equation}
where now
\begin{equation}
  \label{eq:idright}
  \inc{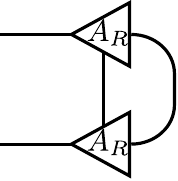} = \hat{1},
\end{equation}
and $D_{L}\le \tilde{D} \le dD_{R}$.

By always taking \textit{thin} QRs we can put the MPS into canonical form by iteratively applying \eq{rightfac} from the
right and \eq{leftfac} from the left and contracting $C$ into the neighboring site

\begin{equation}
  \label{eq:mpsmixed}
  \inc{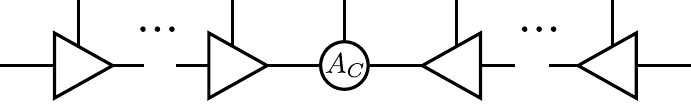}.
\end{equation}
\noindent
In doing so one will always be left with one site that is not of the form $A_{L}$ or $A_{R}$. This site will be known as
the orthogonality center and will be denoted $A_{C}$. The orthogonality center may be placed on any site of the MPS. If
$A_{C}$ is on site $1$($N$) the MPS is said to be in right(left)-canonical form, while its being on any other site is
known as mixed-canonical form.

One may also gauge the MPS such that $C$ lies between two sites
\begin{equation}
  \label{eq:mpsC}
  \inc{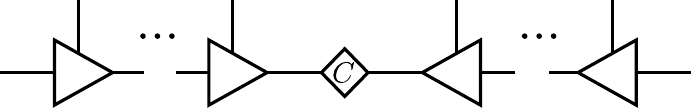}.
\end{equation}

\section{1TDVP with increasing bond-dimensions}
\label{sec:tdvp}

\noindent
The tangent space projector appearing in \eq{tdvp} can be split up in the following manner \cite{lubich_time_2015}

\begin{multline}
  \label{eq:proj}
  \proj(\{\tilde{D}\}) = \sum_{i=1}^{N}\hat{P}_{A_{C}}^{(i)}(\tilde{D}_{i-1},\tilde{D}_{i})-\sum_{i=1}^{N-1}\hat{P}_{C}^{(i)}(\tilde{D}_{i}),
\end{multline}
where
\begin{equation}
  \label{eq:Pac}
  \hat{P}_{A_{C}}^{(i)}(\tilde{D}_{i-1},\tilde{D}_{i}) = \cdots \inc[1.3]{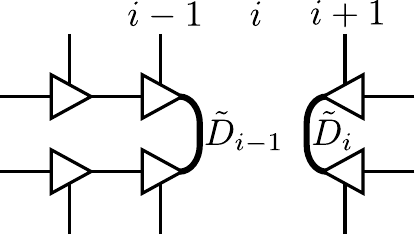} \cdots,
\end{equation}
and
\begin{equation}
  \label{eq:Pc}
  \hat{P}_{C}^{(i)}(\tilde{D}_{i}) = \cdots \inc[1.3]{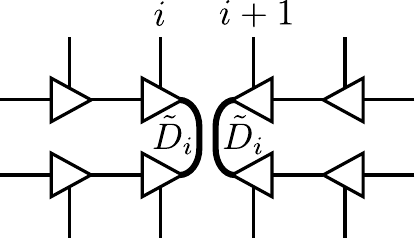} \cdots.
\end{equation}
\noindent
We have explicitly included the dependence on the bond-dimension brought about by expanding the sub-spaces of the MPS
site tensors as described in the previous section. We note that each term in the first sum of \eq{proj} is only affected
by expanding the sites $i-1$ and $i+1$ and that each term in the second sum is affected only by expanding the sites $i$
and $i+1$. In this way the terms of the first sum are each dependent on two bond-dimensions ($D_{i}$ and $D_{i-1}$) while
the terms of the second sum are each dependent on one bond-dimension $(D_{i})$. We may use \eq{proj} to project the MPS
onto a manifold of increased bond-dimension $\{\tilde{D}\}$, while selecting $\{\tilde{D}\} = \{D\}$ will reduce
\eq{proj} to the ordinary fixed bond-dimension projector of 1TDVP.

The reason for splitting the projector in this manner is that, on substituting \eq{proj} into \eq{tdvp}, each term may
be integrated exactly. For example, by gauging the MPS as in \eq{mpsmixed} with $A_{C}$ on site $i$ the operator
$\hat{P}_{A_{C}}^{(i)}\hat{H}$ affects only site $A_{C}$ and so may be written as an effective Hamiltonian $H(i)$ which acts
on this site

\begin{align}
  \label{eq:heff}
  H(i) &= \inc[-1.1]{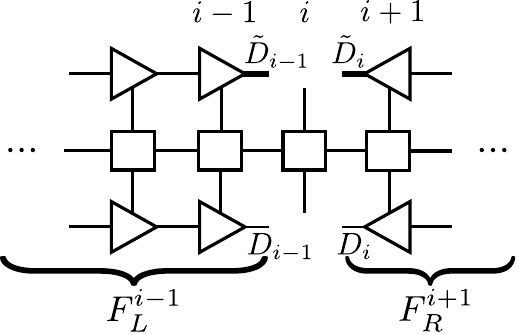} \\
  \label{eq:heff2}
       &= \inc{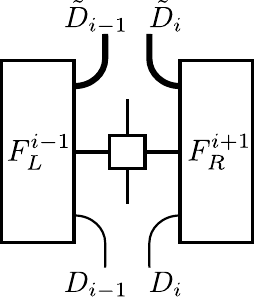}.
\end{align}
Then by making all other sites time-independent we can write the exact evolution of $A_{C}$ as 
\begin{equation}
  \label{eq:Act}
  A_{C}(i,t) = \exp[-iH(i)t]A_{C}(i,0).
\end{equation}

Similarly by writing the MPS as in \eq{mpsC} with $C$ between sites $i$ and $i+1$ and making only $C$ time-dependent we have
\begin{equation}
  \label{eq:Ct}
  C(i,t) = \exp[+iK(i)t]C(i,0),
\end{equation}
with the effective Hamiltonian
\begin{equation}
  \label{eq:keff}
  K(i) = \inc{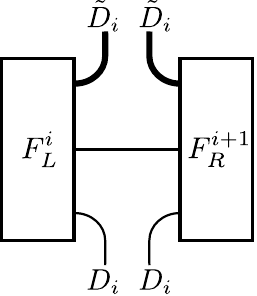}.
\end{equation}
\noindent
With these solutions the entire MPS can be evolved using a Lie-Trotter splitting
\cite{descombes_lietrotter_2013,lubich_quantum_2008,trotter_product_1959} by sweeping from left to right along the chain
and evolving each $A_{C}$ and $C$ by a time step $\Delta t$. If this left to right sweep is composed with a reverse sweep
from right to left then this procedure constitutes a second-order integrator with error $O(\Delta t^{3})$.

With the sub-space expansions employed in \eq{proj} the effective Hamiltonians become capable of increasing the
bond-dimensions, whereas in normal 1TDVP they would leave them unchanged. For example, $H(i)$ takes an MPS site tensor
$A_{C}$ with right and left bond-dimensions $D_{i-1}$ and $D_{i}$ respectively and outputs a tensor with bond-dimensions
$\tilde{D}_{i-1}$ and $\tilde{D}_{i}$.

Using the effective Hamiltonians of \eqs{heff}{keff} thus allows us to grow our MPS to accommodate increasing
entanglement.

We emphasize here that, whenever any of the bond-dimensions are increased, the projector of \eq{proj} is not unique due
to the non-uniqueness of $Q_{2}$ in \eq{qr}. Indeed, we can chose $Q_{2}$ to be any set of states which are orthogonal to
all the states in $Q_{1}$. This means that the projector of \eq{proj} is not necessarily optimal, ie will not find the
best MPS on the increased bond-dimension manifold to approximate the time evolved wave-function. It is possible to
consider the question of optimizing the projector to find the best set of states with which to expand the sub-spaces, as
done in \cite{vanhecke_tangent-space_2020}. Doing so could lead to faster convergence in the bond-dimension although
perhaps at the expense of a more time consuming bond update step. We will leave the pursuit of this question for future
work and here simply use the $Q_{2}$ returned by the QR routine as this requires no additional computational effort to
obtain.

\begin{figure}
  \includegraphics[width=\columnwidth]{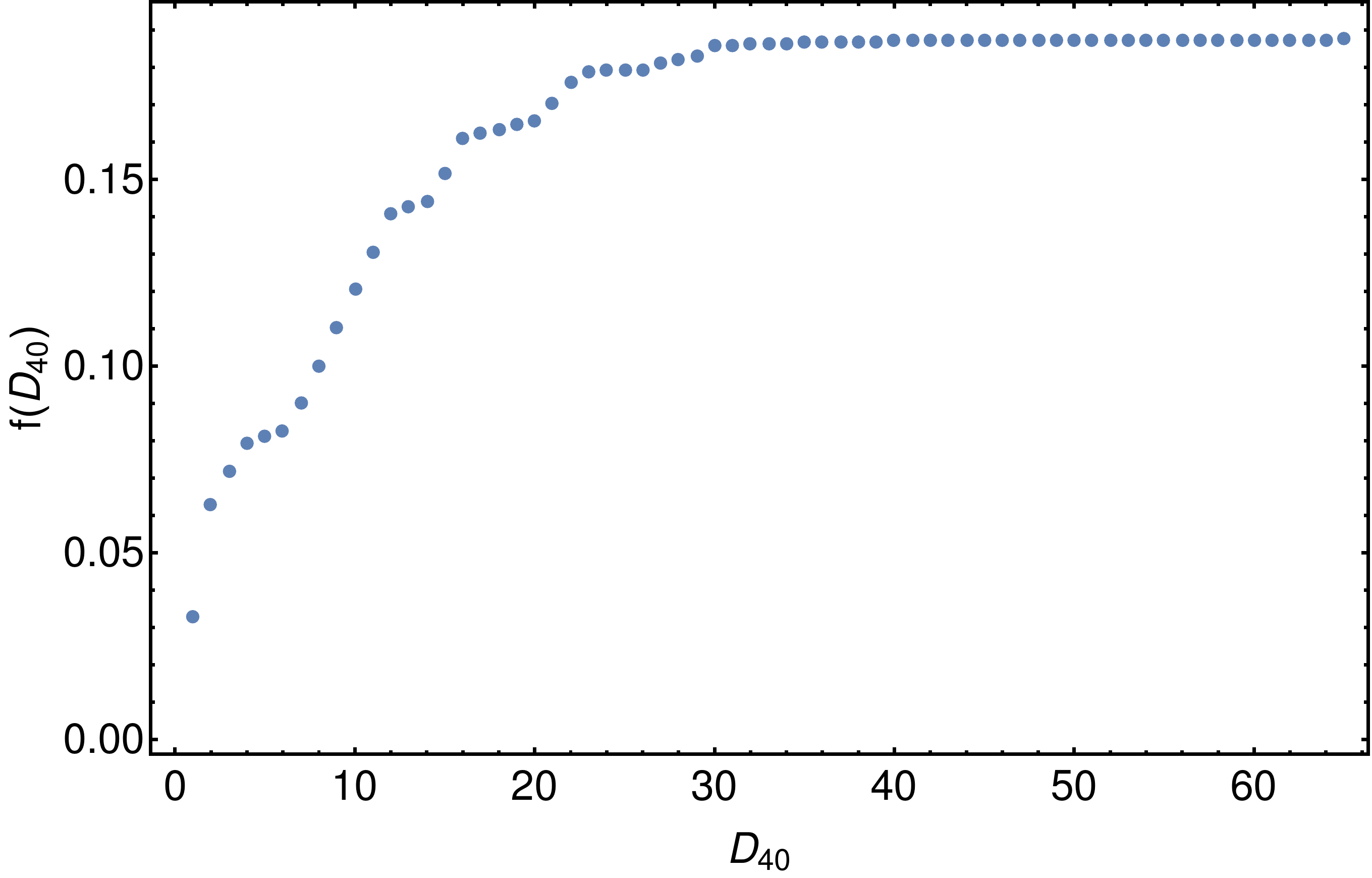}
  \caption{An example of the convergence measure $f$. The $f$ shown is taken from the $p=1.0e-6$ run of the example of
    section \ref{sec:eg} and shows the convergence in $D_{40}$, ie the bond linking the spin to the first site of chain
    $a$, at time $t=10$. The absolute value of $f$ has no meaning, however the fact that it is constant above $D=30$
    tells us that increasing $D_{40}$ above this value will have very little effect on the projection error. The fact
    that the gradient of $f$ is non-monotonic suggests an optimization is possible via the re-ordering of the $Q_{2}$
    states.}
  \label{fig:f}
\end{figure}

In order to use to \eqs{heff}{keff} to dynamically grow the MPS bond-dimensions we require a measure of convergence to
allows us to select appropriate values for $\{\tilde{D}\}$. For this we take the norms of the effective Hamiltonian
applied to their respective MPS site tensors $||H(i)A_{C}(i)||$ and $||K(i)C(i)||$. Since each bond-dimension appears in three terms of the projector we must consider three such
norms for each $D_{i}$. We thus define the following convergence measure for the bond-dimension $D_{i}$

\begin{multline}
  \label{eq:conv}
  f(\tilde{D_{i}}) \equiv ||H(i)A_{C}(i)||^{2} + ||K(i)C(i)||^{2} \\+ ||H(i+1)A_{C}(i+1)||^{2}.
\end{multline}
\noindent
An example of $f$ for a particular bond-dimension is shown in \fig{f}.

By calculating $f$ for the $N-1$ variable bond-dimensions (excluding $D_{0}$ and $D_N$ since these are always fixed at 1)
between each sweep of 1TDVP we are able to determine appropriate bond-dimensions for the next sweep.

The new bond-dimensions are determined, for a precision $p$, as being the smallest $D_{i}$ for which
\begin{equation}
  \label{eq:prec}
  \frac{f(D_{i}+1)}{f(D_{i})}-1 \le p.
\end{equation}

The computational cost of updating the bond-dimensions is as follows. First the tensors $F_{L}^{i}$ for $i\in[1,N-1]$ must
be computed, requiring a left to right QR sweep of the MPS. The tensors $F_{R}^{i}$ will already be available from the
previous right to left sweep of 1TDVP. The overhead of this additional QR sweep may be mitigated by using the $A_{C}$s
produced as a shortcut to computing observables along the chain. Following this the $2N-1$ tensors $H(i)A_{C}(i)$ and
$K(i)C(i)$ can be computed for chosen maximum values of $\{\tilde{D}\}$. This may be done in series or in parallel. Once
all the tensors $H(i)A_{C}(i)$ and $K(i)C(i)$ have been computed, the $N-1$ $f$s may be computed for different values of
$\{\tilde{D}\}$ by simply truncating these tensors and calculating the norms, an operation which carries almost no
additional overhead.

It is clear that this bond-update step will take only a small fraction of the time required for a 1TDVP sweep. In 1TDVP
by far the most expensive operation is the application of the exponentiated effective Hamiltonians of \eqs{Act}{Ct},
carried out using the Krylov method \cite{hochbruck_krylov_1997}, which involves many applications of $H(i)$ and $K(i)$
respectively. In the bond update step these operations are replaced by the calculation of $H(i)A_{C}(i)$ and $K(i)C(i)$
which each require only a single application of $H(i)$ and $K(i)$ respectively. Furthermore, the calculation of these
tensors may be parallelized $(2N-1)$-fold.

\section{Non-equilibrium steady-states in two-temperature open systems}
\label{sec:eg}

\noindent
In this section we demonstrate the utility and power of our method for open quantum systems with numerical simulations of a two-level system strongly and simultaneously coupled to two bosonic
baths at different temperatures. This class of two-environment models has both wide-ranging practical applications -
such as studying heat and charge transfer in nanodevices \cite{benenti_fundamental_2017,dubi_colloquium_2011,dhar_heat_2008}, as
well as fundamental relevance for quantum thermodynamics, decoherence, and non-equilibrium steady states \cite{guo_critical_2012,zhou_symmetry_2015,bruognolo_two-bath_2014,segal_spin-boson_2005,chen_steady_2020}. 

Here we consider two baths, labeled $a$ and $b$, with identical linear couplings and spectral densities, at different inverse
temperatures $\beta_{a}$ and $\beta_{b}$ respectively ($\beta=1/T$). The system-bath Hamiltonian is given by

\begin{equation}
  \label{eq:ham}
  \hat{H} = \frac{\omega_{0}}{2}\sigma_{z}+\hat{H}_{I}^{a}+\hat{H}_{I}^{b}+\hat{H}_{B}^{a}+\hat{H}_{B}^{b},
\end{equation}
where
\begin{align}
  \label{eq:Hia}
  &\hat{H}_{I}^{a} = \sigma_{x}\otimes\sum_{k}(g_{k}^{*}\hat{a}_{k}+g_{k}\hat{a}_{k}^{\dagger})\\
  &\hat{H}_{I}^{b} = \sigma_{x}\otimes\sum_{k}(g_{k}^{*}\hat{b}_{k}+g_{k}\hat{b}_{k}^{\dagger})\\
  &\hat{H}_{B}^{a}=\sum_{k}\omega_{k}\hat{a}_{k}^{\dagger}\hat{a}_{k}\\
  &\hat{H}_{B}^{b}=\sum_{k}\omega_{k}\hat{b}_{k}^{\dagger}\hat{b}_{k}.
\end{align}

\noindent
The bath parameters are defined in terms of the spectral density
$J(\omega)\equiv\pi\sum_{k}|g_{k}|^{2}\delta(\omega-\omega_{k})$ which we take to be Ohmic with a hard cut-off at frequency
$\omega_{c}$
\begin{equation}
  \label{eq:specdens}
  J(\omega) = 2\pi\alpha\omega\theta(\omega_{c}-\omega).
\end{equation}
\noindent
The initial condition $\hat{\rho}(0)$ is taken to be an uncorrelated (product) state of the spin and baths, which - because of the baths' finite temperatures - must be described by a mixed state, i.e. a density matrix 
\begin{equation}
  \label{eq:init}
  \hat{\rho}(0)=\ket{\uparrow_{z}}\bra{\uparrow_{z}}
  \otimes\frac{e^{-\hat{H}_{B}^{a}\beta^{a}}}{\Tr\{e^{-\hat{H}_{B}^{a}\beta^{a}}\}}
  \otimes\frac{e^{-\hat{H}_{B}^{b}\beta^{b}}}{\Tr\{e^{-\hat{H}_{B}^{b}\beta^{b}}\}}.
\end{equation}

Remarkably, despite the initial condition containing two statistically mixed thermal density matrices, it has recently been shown by Tamescello \textit{et al}. that the reduced dynamics of the spin can be obtained from a \emph{single} MPS simulation of a \emph{pure} system-environment wave function \cite{tamascelli_nonperturbative_2018,tamascelli_efficient_2019}. This follows from the formal equivalence of the influence function of a physical environment at inverse temperature $\beta$ with a spectral density $J(\omega)$ and a proxy, zero-temperature environment that is extended over both positive and negative frequencies and which possesses an effective, $\beta$-dependent spectral density $J(\omega,\beta)$ - see Refs. \cite{tamascelli_efficient_2019,
  tamascelli_nonperturbative_2018, tamascelli_supplemental_nodate}. The thermal behavior of the environment is then exactly reproduced using a set of initially empty modes of positive and negative frequencies. This
is possible due to the fact that the system cannot distinguish between the occupation of an initially empty negative
mode and the depletion of an initially occupied positive mode. A similar idea, formulated in terms of two-mode squeezed states, is at the heart of the thermofield approach of Ref.  \cite{de_vega_how_2015}, where one can directly see how the relative couplings to modes at $\pm \omega$ enforce the property of detailed balance contained in the physical correlation function of the original, finite-temperature environment.  

Following the introduction of the effective $J(\omega,\beta)$, it then becomes possible to apply a unitary transformation of the harmonic environment - the so-called chain mapping \cite{chin_exact_2010,de_vega_how_2015} - which maps the two baths onto two
separate $1D$, nearest-neighbour tight-binding chains that couple at one end to the spin. The resulting transformed system, which is formally equivalent to the initial open system problem, is shown in Fig. \ref{fig:bd}  The hopping parameters and site energies of the bosonic chains are
determined completely by the effective spectral density $J(\omega,\beta)$.

We denote the transformed bath operators as
\begin{equation}
  \label{eq:Ua}
  \hat{A}_{n}=\sum_{k}U_{kn}(\beta_{a})\hat{a}_{k}
\end{equation}
\begin{equation}
    \label{eq:Ub}
    \hat{B}_{n}=\sum_{k}U_{kn}(\beta_{b})\hat{b}_{k},
\end{equation}
\noindent
and our transformed initial condition is
\begin{equation}
  \label{eq:mpsprod}
  \ket{\psi} = \ket{0}_{a}\otimes\ket{\uparrow_{z}}\otimes\ket{0}_{b},
\end{equation}
where $\ket{0}_{a(b)}$ is the vacuum state of bath $a(b)$. Physically, such an initial condition could correspond to the sudden connection of two reservoirs by a qubit or few-state nanoscopic junction. Since $\ket{\psi}$ is a product state, it can be represented
as an MPS with all bond-dimensions equal to 1. Under conventional 1TDVP it would not be possible to use such an initial
MPS since the bond-dimensions would be fixed at 1 throughout the whole simulation, thus only capturing the on-site
evolution. However, with our bond-adaptive 1TDVP method we are able to start from this extremely simple initial condition and the increase the bond-dimensions on the fly to capture the
growing entanglement, as, when and where it is needed.

\subsection{Numerical results and observations of transient heat flows}

\begin{figure}
  \includegraphics[width=\columnwidth]{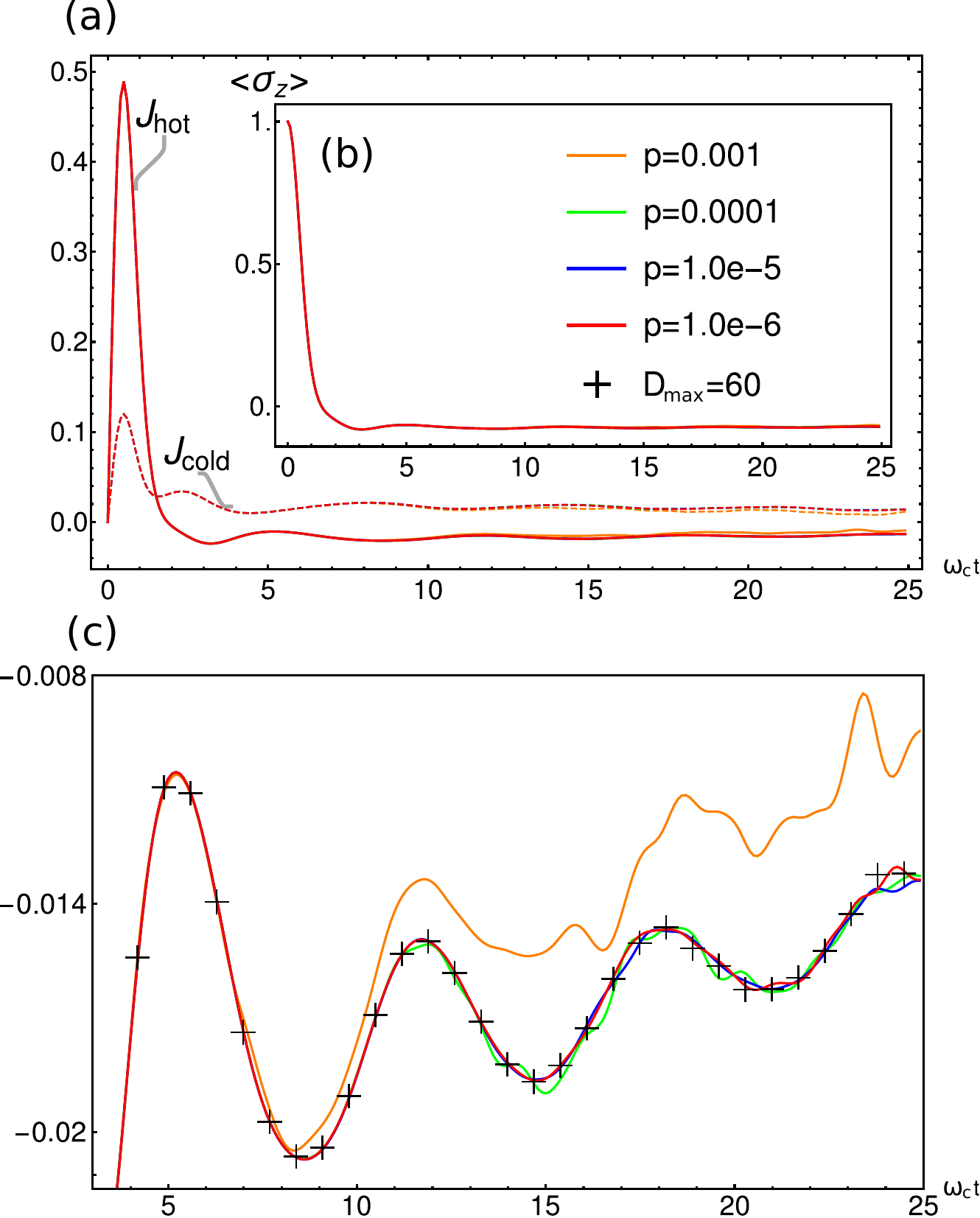}
  \caption{Dynamics of a two level system with splitting $\omega_{0}=0.2\omega_{c}$ coupled between two harmonic baths,
    temperatures $\beta_{a}=100/\omega_{c}$, $\beta_{b}=1/\omega_{c}$, with Ohmic spectral densities, coupling strength
    $\alpha=0.2$. The Foch spaces of the bath modes were truncated to $15$ states. Results were obtained using 1TDVP
    with dynamic bond-dimensions for 4 different values of the precision $p$ and $D_{lim}=60$ and checked against
    conventional 1TDVP with $D_{max}=60$. (a) $\langle\hat{J}_{a}(t)\rangle$ (dashed line) and
    $\langle\hat{J}_{b}(t)\rangle$ (solid line). (b) $\langle\sigma_{z}(t)\rangle$
    ($\langle\sigma_{x}(t)\rangle=\langle\sigma_{y}(t)\rangle=0$). (c) Expanded view of $\langle\hat{J}_{b}(t)\rangle$,
    black crosses show results obtained from fixed bond-dimension 1TDVP.}
  \label{fig:Jsz}
\end{figure}

\begin{figure}
  \includegraphics[width=\columnwidth]{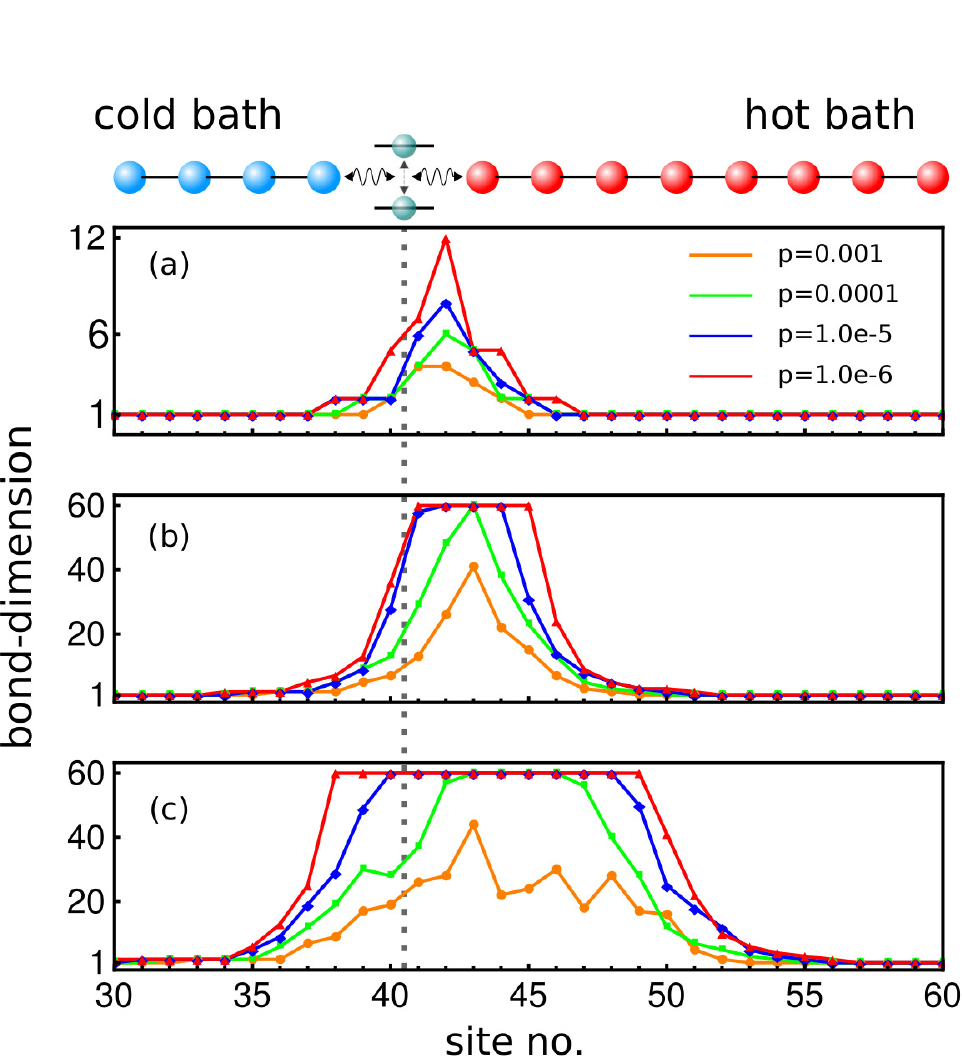}
  \caption{Bond-dimensions at times $\omega_{c}t=1$ (a), $\omega_{c}t=5$ (b) and $\omega_{c}t=10$ (c). The spin is on site 41 of the chain. The
    bond-dimensions grow faster and expand away from the spin quicker for the hot bath than for the cold bath, reflecting
    the different computational resources necessary to simulate them.}
  \label{fig:bd}
\end{figure}

We run the dynamics with four values of the precision $p$ to show convergence. For simplicity and speed we set the local limit $D_{lim}=60$. For comparison we also ran the simulation with conventional 1TDVP with a fixed
maximum bond-dimension $D_{max}=60$. Setting $D_{lim}=60$ guarantees that the dynamic bond-dimension simulations will all
be faster than normal 1TDVP at $D_{max}=60$. The Foch spaces of the bath modes are truncated to $15$ states \cite{woods_simulating_2015}. The simulations were run on an Intel\textsuperscript{\textregistered} Xeon\textsuperscript{\textregistered} W-2123
Processor. We find that a significant speed up is achieved with the most precise dynamic-bond simulation ($p=1.0e-6$)
completing in 1 hour 26 minutes versus 5 hours 55 minutes for the fixed-bond run.

The spin dynamics are shown in \fig{Jsz} for the following choice of parameters in units of $\omega_{c}$:
$\beta_{a}=100/\omega_{c}, \beta_{b}=1/\omega_{c}, \alpha=0.2$ and $\omega_{0}=0.2\omega_{c}$. The chosen value of $\alpha$ is considered non-perturbative
for Ohmic baths \cite{weiss2012quantum}. The spin evolves from its initial pure state of up along $\ket{\uparrow_{z}}$ into a statistical mixture of up and down
with a slight predominance for spin down in the steady state, due to the presence of the cold environment (the hot environment has a temperature x20 greater than the spin energy level splitting) (\fig{Jsz}(b)).

Another interesting dynamical observable is the heat flow between the spin and the two baths. We define the following
operators
\begin{equation}
  \label{eq:jcold}
  \hat{J}_{a}=\sigma_{y}\otimes(\hat{A}_{0}^{\dagger}+\hat{A}_{0}),
\end{equation}
and
\begin{equation}
  \label{eq:jhot}
  \hat{J}_{b}=\sigma_{y}\otimes(\hat{B}_{0}^{\dagger}+\hat{B}_{0}),
\end{equation}
which measure the heat flux from the spin to baths $a$ and $b$ respectively. At early times we see from \fig{Jsz}(a)
that heat flows out of the spin into both hot and cold baths, since the spin begins in an inverted state, and that later
a steady state is established in which net heat flows into the spin from the hot bath ($b$) and out of the spin into the
cold bath ($a$). We see also from the expanded view of $\hat{J}_{b}$ in \fig{Jsz}(c) that these results are well
converged w.r.t. the results of the fixed bond-dimension simulation, clearly demonstrating the advantage
of our approach.

The bond-dimensions for all four values of $p$ at three different time steps are also shown in \fig{bd}. We observe that the simulation of the chains becomes more expensive for higher temperatures, and we have traced back to the properties of the chain parameters and effective thermal spectral densities: stronger intra-chain coupling for higher temperature environments causes entanglement between the sites to grow faster and causes excitations to travel faster along the chain. Longer chains are thus required in order to avoid boundary effects, and larger bond dimensions are needed for the stronger quantum correlations across the chain sites. We see from
\fig{bd} that the new method is able to account for both of these effects when optimizing the bond-dimensions by giving
larger bond-dimensions to chain $b$ which propagate outwards faster to match the progress of the perturbation along the
chain.

Indeed, we find that this new approach is particularly effective for this class of open system problems, in which the
environment is modeled with a chain initialized in the vacuum state. In particular, little thought now has to be paid to
the choice of chain lengths since any chain sites not touched by the perturbation will remain with trivial
bond-dimensions and so the updates on these sites will take almost no time. Under fixed bond-dimension 1TDVP, however
the chain lengths have a significant impact on simulation time and so must be chosen carefully
\cite{tamascelli_supplemental_nodate}. As a very large number of open quantum system problems involve 'impurity' like dynamics of a spatially localised system with an initiialy uncorrelated environment state, taking full advantage of the evolving range of correlations in an automated way could lead to very efficient `black-box' algorithms for open system problems. This idea may be straight forwardly extended to tree-MPS which will allow the simulation of systems with
complex multi-environment interactions, whereupon, the advantages demonstrated here will
become even more important and could be further combined with the `entanglement renormalisation techniques` introduced in Ref.  \cite{schroder_tensor_2019} for molecular open quantum dynamics. 

Finally, analysis of the distribution and development of correlations, as in Fig. \ref{fig:bd}, can also provide interesting physical insights into the underlying microscopic dynamics, particularly when considering non-equilibrium problems where quantum systems interact with multiple quantum environments, i.e. in nanoscale energy harvesting, transport and sensing. We believe our adaptive 1TDVP could prove very useful in such problems, and perhaps several others types that lie outside of the typical scope of open quantum systems.

\noindent

\bibliographystyle{h-physrev}
\bibliography{dynamic1tdvp}

\end{document}